\documentclass[12pt,reqno]{amsart}

\usepackage{amsmath,amsthm,amssymb,comment,fullpage}
\usepackage{times}
\usepackage[T1]{fontenc}
\usepackage{mathrsfs}
\usepackage{latexsym}
\usepackage[dvips]{graphics}
\usepackage{epsfig}
\usepackage{accents}
\usepackage{float}
\usepackage{amsmath,amsfonts,amsthm,amssymb,amscd}
\input amssym.def
\input amssym.tex
\usepackage{color}
\usepackage{hyperref}
\usepackage{url}
\usepackage{breakurl}
\usepackage{comment}
\usepackage{mathtools}
\newcommand{\bburl}[1]{\textcolor{blue}{\url{#1}}}

\newcommand{\E}{\mathbb{E}}
\renewcommand{\E}{\mathbb{E}}

\numberwithin{equation}{section}

\newtheorem{thm}{Theorem}[section]

\theoremstyle{plain}

\newtheorem{rek}[thm]{Remark}

\newcommand\be{\begin{equation}}
\newcommand\ee{\end{equation}}
\newcommand\bea{\begin{eqnarray}}
\newcommand\eea{\end{eqnarray}}
\newcommand\bi{\begin{itemize}}
	\newcommand\ei{\end{itemize}}
\newcommand\ben{\begin{enumerate}}
	\newcommand\een{\end{enumerate}}
\newcommand\bc{\begin{center}}
	\newcommand\ec{\end{center}}
\newcommand\ba{\begin{array}}
	\newcommand\ea{\end{array}}

\newcommand{\ncr}[2]{{#1 \choose #2}}

\newcommand{\hr}[1]{\href{#1}{\url{#1}}}

\newcommand{\vectwo}[2]
{\left(\begin{array}{c}
                        #1    \\
                        #2
                          \end{array}\right) }

\newcommand{\mattwo}[4]
{\left(\begin{array}{cc}
		#1  & #2   \\
		#3 &  #4
	\end{array}\right) }

\newcommand*{\reff}[1]{\hyperref[#1]{\ref{#1}}}

\title{Lessons from the German Tank Problem}

\author{George Clark}
\email{\textcolor{blue}{\href{mailto:gpclark96@gmail.com}{gpclark96@gmail.com}}}
\address{Department of Mathematics and Statistics, Williams College, Williamstown, MA 01267}

\author{Alex Gonye}
\email{\textcolor{blue}{\href{mailto:aag2@williams.edu}{aag2@williams.edu}}}
\address{Department of Mathematics and Statistics, Williams College, Williamstown, MA 01267}

\author{Steven J. Miller}
\email{\textcolor{blue}{\href{mailto:sjm1@williams.edu, Steven.Miller.MC.96@aya.yale.edu}{sjm1@williams.edu,Steven.Miller.MC.96@aya.yale.edu}}}
\address{Department of Mathematics and Statistics, Williams College, Williamstown, MA 01267}
\curraddr{Department of Mathematical Sciences, Carnegie Mellon University, Pittsburgh, PA 15213}

\thanks{The third named author was partially supported by NSF Grant DMS1561945. Parts of this paper were given by the first two named authors as their senior colloquium; we thank our colleagues in the department for comments. The third author has presented this topic at several conferences and programs, and thanks the participants for many suggestions; for video see \bburl{https://www.youtube.com/watch?v=quV-MCB8Ozs} or \bburl{https://youtu.be/I3ngtIYjw3w}.}

\subjclass[2010]{62J05, 60C05 (primary), 05A10 (secondary)}

\keywords{German Tank Problem, Binomial Identities, Regression}

\date{\today}

\begin{document}

\begin{abstract} During World War II the German army used tanks to devastating advantage. The Allies needed accurate estimates of their tank production and deployment. They used two approaches to find these values: spies, and statistics. This note describes the statistical approach. Assuming the tanks are labeled consecutively starting at 1, if we observe $k$ serial numbers from an unknown number $N$ of tanks, with the maximum observed value $m$, then the best estimate for $N$ is $m(1 + 1/k) - 1$. This is now known as the German Tank Problem, and is a terrific example of the applicability of mathematics and statistics in the real world. The first part of the paper reproduces known results, specifically deriving this estimate and comparing its effectiveness to that of the spies. The second part presents a result we have not found in print elsewhere, the generalization to the case where the smallest value is not necessarily 1. We emphasize in detail why we are able to obtain such clean, closed-form expressions for the estimates, and conclude with an appendix highlighting how to use this problem to teach regression and how statistics can help us find functional relationships.  \end{abstract}

\thanks{We thank our colleagues at Williams and audiences at several talks over the years, and the referee, for valuable feedback that improved the exposition, and Jason Zhou for pointing out some typos in an earlier draft.}

\maketitle
\tableofcontents

%%%%%%%%%%%%%%%%%%%%%%%%%%%%%%%%%%%%%%%%%%%%%%%%%%%%%%%%%%%%%%%%%%%%%%%%%%%%%%%%%%%%%%%%%%%%%%%%%%%%%%%%%%%%%%%%%%%%%%%
%%%%%%%%%%%%%%%%%%%%%%%%%%%%%%%%%%%%%%%%%%%%%%%%%%%%%%%%%%%%%%%%%%%%%%%%%%%%%%%%%%%%%%%%%%%%%%%%%%%%%%%%%%%%%%%%%%%%%%%
%%%%%%%%%%%%%%%%%%%%%%%%%%%%%%%%%%%%%%%%%%%%%%%%%%%%%%%%%%%%%%%%%%%%%%%%%%%%%%%%%%%%%%%%%%%%%%%%%%%%%%%%%%%%%%%%%%%%%%%
\section{Introduction}

In this paper we revisit a famous and historically important problem which has since become a staple in many probability and statistics classes: the German Tank Problem. This case study illustrates that one does not need to use the most advanced mathematics to have a tremendous impact on real world problems; the challenge is frequently in creatively using what one knows.\footnote{Another great example is the famous Battle of Midway and the role the cryptographers played in figuring out the Japanese target; see for example \cite{A0}.}

\begin{figure}
\begin{center}
\scalebox{.75}{\includegraphics{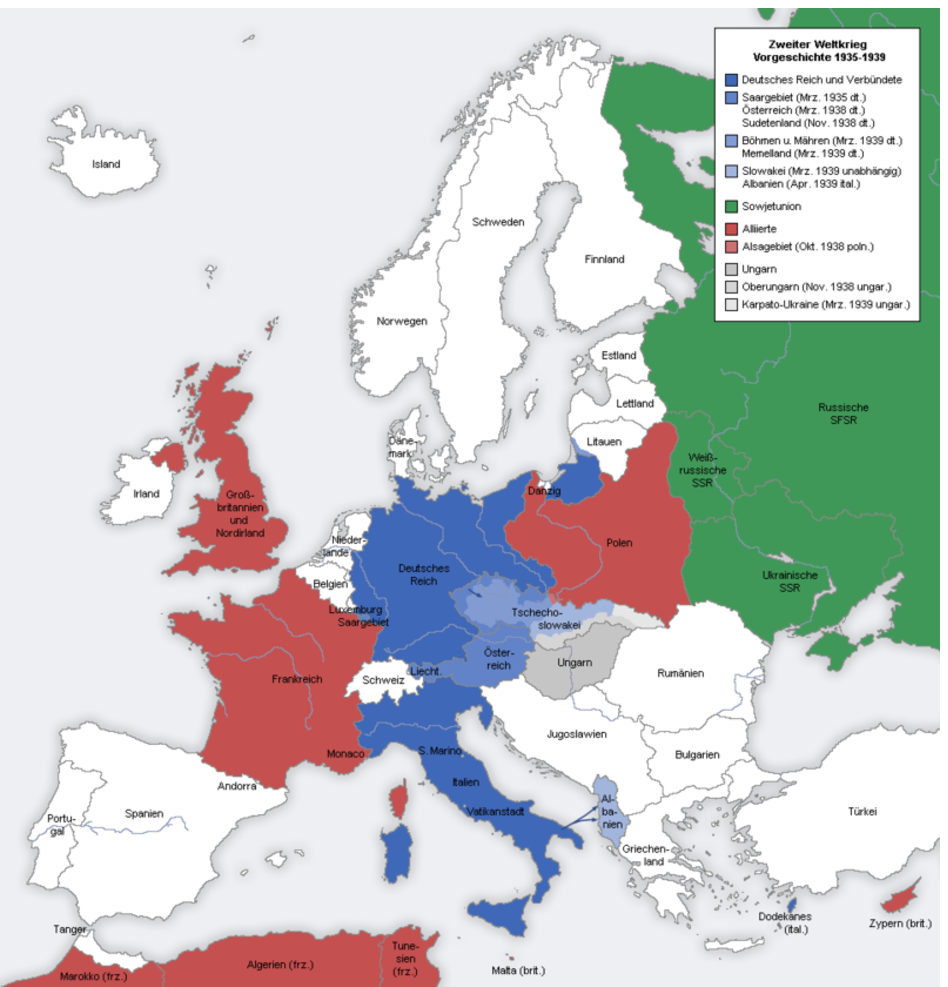}}\ \ \scalebox{.75}{\includegraphics{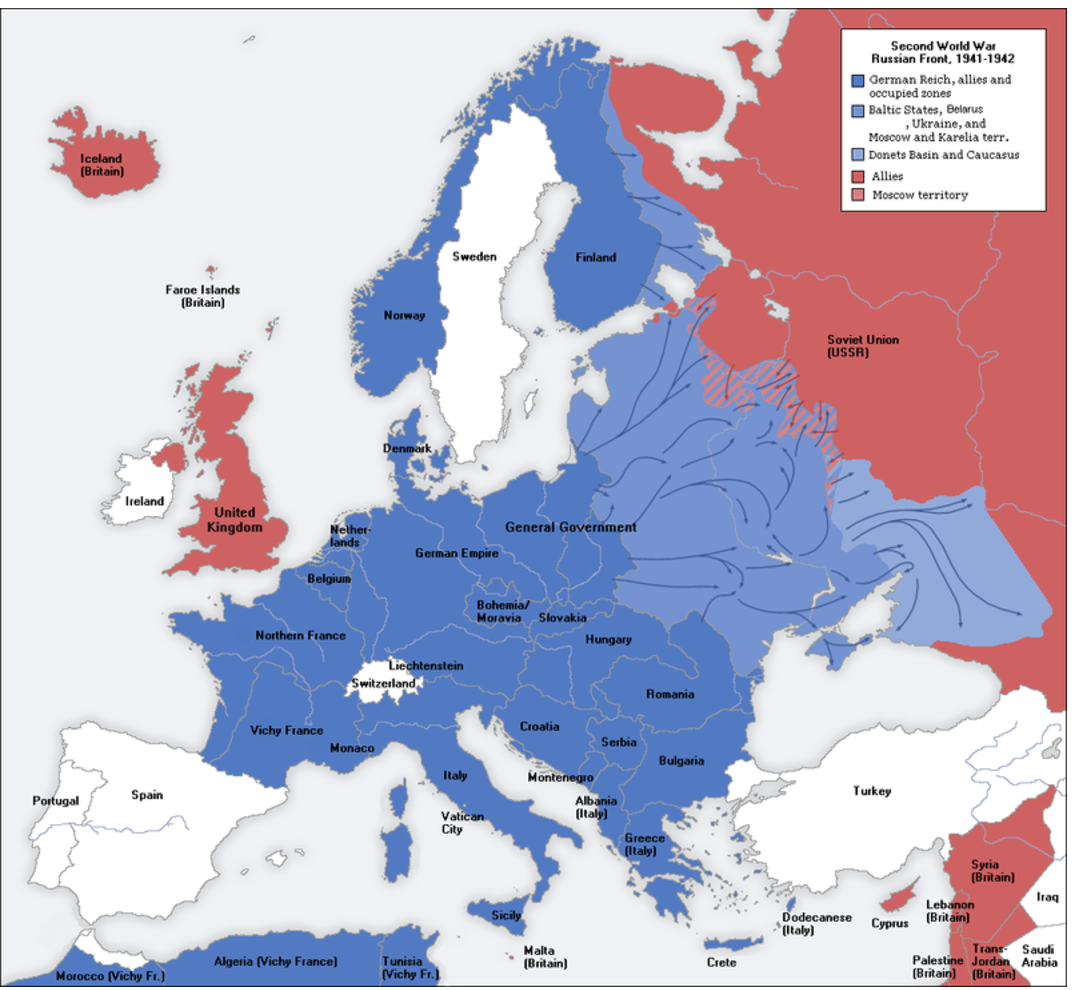}}\caption{\label{fig:maps} Left: Europe before the start of major hostilities. Right: Europe in 1942. Images from Wikimedia Commons from author San Jose.}
\end{center}\end{figure}

By the end of 1941 most of continental Europe had fallen to Nazi Germany and the other Axis powers, and by 1942 their forces had begun their significant advance in the Eastern Front deep into the Soviet Union; see Figure \ref{fig:maps} for an illustration of their rapid progress, or \bburl{https://www.youtube.com/watch?v=WOVEy1tC7nk} for an animation of territorial gains day by day. A key component to their rapid conquests lay with their revolutionary use of tanks in modern warfare. While other militaries, most notably France, used tanks as a modern, armored form of cavalry, the Germans were the first to fully utilize tanks' speed and strength to their advantage. Tanks would move rapidly and punch through enemy lines creating gaps which German infantry would stream through. Once through the holes in the line, the Germans would wreak havoc on lines of communication, creating logistical nightmares for the combatants left on the front lines. This lightning fast warfare has become dubbed Blitzkrieg (or lighting war) by many historians.

With the Nazis utilizing tanks with such devastating results, it was essential for the Allies to stop them. A key component to the solution was figuring out how many tanks the Germans were building, or had deployed in various theaters, in order to properly allocate resources. As expected, they tried spying (both with agents and through decrypting intercepted messages) to accurately estimate these numbers. It was essential that the Allies obtain accurate values, as there is a tremendous danger in both under- and over-estimating the enemy's strength. The consequence of underestimating is clear, as one could suddenly be outnumbered in battle. Overestimating is also bad, as it can lead to undo caution and failure to exploit advantages, or to committing too many resources in one theater and thus not having enough elsewhere. The U.S. Civil War provides a terrific example of these consequences, where General George McClellan would not have his Union army take the field against the Confederates because, in his estimation, his forces were greatly outnumbered though in fact they were not. This situation led to one of President Abraham Lincoln's many great quips: \emph{If General McClellan isn't going to use his army, I'd like to borrow it for a time.}\footnote{There are many different phrasings of this remark; this one is taken from \bburl{ https://thehistoriansmanifesto.wordpress.com/2013/05/13/best-abraham-lincoln-quotes/}.} Considering how close Pickett's charge came to succeeding at Gettysburg, or what would have happened if Sherman hadn't taken Atlanta before the 1864 elections (where McClellan, now as the Democrat nominee for president, was running against Lincoln on a platform of a negotiated peace), the paralysis from incorrect analysis could have changed the outcome of the war.

Returning to World War II and the problem of determining the number of tanks produced and facing them in the field, the Allies understandably wanted to have a way to evaluate the effectiveness of their estimates. During a battle, they realized that the destroyed and captured tanks had serial numbers on their gearboxes which could help with this problem. Assuming that the serial numbers are sequential and start with 1, given a string of observed numbers one can try to estimate the largest value. This discovery contributed to the birth of the statistical method, the use of a population maximum formula, and changed both the war and science.\footnote{There are advantages in consecutive labeling; for example, it simplifies some maintenance issues as it is clear which of two tanks is older.}

The original variant of the problem assumes the first tank is numbered 1, there are an unknown number $N$ produced (or in theater), that the numbers are consecutive, and that $k$ values are observed with the largest being $m$. Given this information, the goal is to find $\widehat{N}$, the best estimate of $N$. The formula they derived is $$\widehat{N}\ =\ m\left(1+\frac1{k}\right) - 1.$$

For the reader unfamiliar with this subject, we deliberately do not state here in the introduction how well this formula does versus what was done by spies and espionage.\footnote{That said, as this paper is appearing in a mathematics journal and not a bulletin of a spy agency, we invite the reader to conjecture which method did better.} While this story has been well told before (see for example \cite{A2, A3, A4, A5}), our contribution is to extend the analysis to consider the more general case, namely what happens when we do not know the smallest value. To our knowledge this result has not been isolated in the literature; we derive in \S\ref{sec:unknownminimum} that if $s$ is the spread between the smallest and the largest of the observed $k$ serial numbers, then $$\widehat{N} \ = \ s\left(1 + \frac{2}{k-1}\right) - 1.$$ In Appendix \ref{sec:gtpregression} we show how to use regression to show that these are reasonable formulas, and thus the German Tank Problem can also be used to introduce some problems and subtleties in regression analysis, as well as serve as an introduction to mathematical modeling.

As it is rare to have a clean, closed-form expression such as the ones above, we briefly remark on our fortune. The key observation is that we have a combinatorial problem where certain binomial identities are available, and these lead to tremendous simplifications.

%Each tank had a serial number on the gearbox. These numbers were in ascending order with known minimum (the serial numbers started at 1). Thus the question becomes, based on how many serial numbers observed from destroyed tanks, what is the total population in tanks? Or in other words, what is the number of tanks that Germany is producing each month. We will denote the number of captured German tanks as k; let m be the largest serial number of those capture tanks, and our ultimate goal will be to estimate N the total population of tanks (which is equivalent to the monthly production of tanks). To complete this estimation, we will first need to compute the probability of observing k tanks with maximum m out of a total of N tanks. From there we can figure out the expected value of of observing maximum m and solve for N.

%%%%%%%%%%%%%%%%%%%%%%%%%%%%%%%%%%%%%%%%%%%%%%%%%%%%%%%%%%%%%%%%%%%%%%%%%%%%%%%%%%%%%%%%%%%%%%%%%%%%%%%%%%%%%%%%%%%%%%%
%%%%%%%%%%%%%%%%%%%%%%%%%%%%%%%%%%%%%%%%%%%%%%%%%%%%%%%%%%%%%%%%%%%%%%%%%%%%%%%%%%%%%%%%%%%%%%%%%%%%%%%%%%%%%%%%%%%%%%%
%%%%%%%%%%%%%%%%%%%%%%%%%%%%%%%%%%%%%%%%%%%%%%%%%%%%%%%%%%%%%%%%%%%%%%%%%%%%%%%%%%%%%%%%%%%%%%%%%%%%%%%%%%%%%%%%%%%%%%%
\section{Derivation with a known minimum}\label{sec:knownminimum}

%The formula is very simple and easy to use when we assume the tanks are labeled consecutively starting with 1. If we observe the serial numbers of $k$ tanks, with the largest seen being $m$, then our best guess for the number of tanks is

In this section we prove $$\widehat{N}\ =\ m\left(1+\frac1{k}\right) - 1$$ when we observe $k$ tanks, the largest labeled $m$, and knowing that the smallest number is 1 and the tanks are consecutively numbered. Before proving it, as a smell test we look at some extreme cases. First, we never obtain an estimate that is less than the largest observed number. Second, if there are many tanks and we observe just one (so $k=1$), then $\widehat{N}$ is approximately $2m$. This is very reasonable, and essentially just means that if we only have one data point, it's a good guess that it was in the middle. Further, as $k$ increases the amount we must inflate our observed maximum value decreases. For example, when $k=2$ we inflate $m$ by approximately a factor of $3/2$, or in other words this is saying our observed maximum value is probably about two-thirds of the true value. Finally, if $k$ equals the number of tanks $N$, then $m$ must also equal $N$, and the formula simplifies to $\widehat{N} = N$.

We break the proof into two parts. While we are fortunate in that we are able to obtain a closed-form expression, if we have a good guess as to the relationship we can use statistics to test its reasonableness; we do that in Appendix \ref{sec:gtpregression}. For the proof we first determine the probability that the observed largest value is $m$. Next we compute the expected value, and show how to pass from that to an estimate for $N$. We need two combinatorial results.\\ \

The first is Pascal's identity: \begin{eqnarray}\label{eq:pascalidentity} \ncr{n+1}{r}\ =\ \ncr{n}{r} + \ncr{n}{r-1}.\end{eqnarray} There are many approaches to proving this; the easiest is to interpret both sides as two different ways of counting how many ways we can choose $r$ people from a group of $n+1$ people, where exactly $n$ of these people are in one set and exactly one person is in another set. It is easier to see if we rewrite it as $$\ncr{n+1}{r}\ =\ \ncr{1}{0} \ncr{n}{r} \ +\  \ncr{1}{1}\ncr{n}{r-1};$$ this is permissible because $\ncr{1}{0} = \ncr{1}{1} = 1$. Note the left side is choosing $r$ people from the combined group of $n+1$ people, while the right is choosing $r$ people with the first summand corresponds to not choosing the person from the group with just one person, and the second summand to requiring we chose that person. \hfill $\Box$ \ \\

The second identity involves sums of binomial coefficients: \begin{eqnarray}\label{eq:secondcombidentity}\sum_{m = k}^N \ncr{m}{k}\ =\ \ncr{N+1}{k+1}. \end{eqnarray} We can prove this by induction on $N$, noting that $k$ is fixed. The base case is readily established. Letting $N = k$, we find $$\sum_{m = k}^N \ncr{m}{k}\ = \ \sum_{m = k}^k \ncr{m}{k}\ =\ \ncr{k}{k}\ =\ 1\ =\ \ncr{k+1}{k+1}.$$
For the inductive step, we assume  $$\sum_{m = k}^N \ncr{m}{k}\ =\ \ncr{N+1}{k+1}.$$ Then \begin{eqnarray}\sum_{m = k}^{N+1} \ncr{m}{k} & \ =\ & \left(\sum_{m = k}^N \ncr{m}{k}\right) + \ncr{N+1}{k} \nonumber\\
& \ = \ & \ncr{N+1}{k+1} + \ncr{N+1}{k}\nonumber\\ &\ =\ & \ncr{N+2}{k+1}, \nonumber \end{eqnarray} where the last equality follows from Pascal's identity, \eqref{eq:pascalidentity}. This completes the proof. \hfill $\Box$

While this identity suffices for the original formulation of the German Tank Problem, when we do not know the starting serial number the combinatorics become slightly more involved, and we need a straightforward generalization: \begin{equation}\label{eq:secondcombidentitygeneralized}\sum_{\ell=a}^{b} \ncr{\ell}{a}\ =\ \ncr{b+1}{a+1};\end{equation} the proof follows similarly.

%%%%%%%%%%%%%%%%%%%%%%%%%%%%%%%%%%%%%%%%%%%%%%%%%%%%%%%%%%%%%%
%%%%%%%%%%%%%%%%%%%%%%%%%%%%%%%%%%%%%%%%%%%%%%%%%%%%%%%%%%%%%%
%%%%%%%%%%%%%%%%%%%%%%%%%%%%%%%%%%%%%%%%%%%%%%%%%%%%%%%%%%%%%%
\subsection{The probability that the sample maximum is $m$}

Let $M$ be the random variable for the maximum number observed, and let $m$ be the value we see. Note that there is zero probability of observing a value smaller than $k$ or larger than $N$. We claim for $k \le m \le N$ that $${\rm Prob}(M = m) \ = \ \frac{\ncr{m}{k} - \ncr{m-1}{k}}{\ncr{N}{k}}\ =\ \frac{\ncr{m-1}{k-1}}{\ncr{N}{k}}.$$

We give two proofs. The first is to note there are $\ncr{N}{k}$ ways to choose $k$ numbers from $N$ when order does not matter. The probability that the largest observed is exactly $m$ equals the probability the largest is at most $m$ minus the probability the largest is at most $m-1$.  The first probability is just $\ncr{m}{k} / \ncr{N}{k}$, as if the largest value is at most $m$ then all $k$ observed numbers must be taken from $\{1, 2, \dots, m\}$. A similar argument gives the second probability is $\ncr{m-1}{k} / \ncr{N}{k}$, and the claim now follows by using Pascal's identity to simplify the difference of the binomial coefficients.

We could also argue as follows. If the largest is $m$ then we have to choose that serial number, and now we must choose $k-1$ tanks from the $m-1$ smaller values; thus we find the probability is just $\ncr{m-1}{k-1}/\ncr{N}{k}$.  \hfill $\Box$

\begin{rek} Interestingly, we can use the two equivalent arguments above as yet another way to prove the Pascal identity. \end{rek}

\ \\
%%%%%%%%%%%%%%%%%%%%%%%%%%%%%%%%%%%%%%%%%%%%%%%%%%%%%%%%%%%%%%
%%%%%%%%%%%%%%%%%%%%%%%%%%%%%%%%%%%%%%%%%%%%%%%%%%%%%%%%%%%%%%
%%%%%%%%%%%%%%%%%%%%%%%%%%%%%%%%%%%%%%%%%%%%%%%%%%%%%%%%%%%%%%
\subsection{The best guess for $\widehat{N}$}

We now compute the best guess for $N$ by first finding the expected value of $M$. Recall the expected value of a random variable $M$ is the sum of all the possible values of $M$ times the probability of observing that value. We write $\E[M]$ for this quantity, and thus we must compute $$\E[M] \ := \ \sum_{m=k}^N m {\rm Prob}(M=m)$$ (note we only need to worry about $m$ in this range, as for all other $m$ the probability is zero and thus does not contribute). Once we find a formula for $\E[M]$ we will convert that to one for the expected number of tanks.

Our first step is to substitute in the probability that $M$ equals $m$, obtaining $$\E[M] \ = \  \sum_{m = k}^N m \frac{\ncr{m-1}{k-1}}{\ncr{N}{k}}.$$ Fortunately this sum can be simplified into a nice closed-form expression; it is this simplification that allows us to obtain a simple formula for $\widehat{N}$. We expand the binomial coefficients in the expression for $\E[M]$ and then use our second combinatorial identity, \eqref{eq:secondcombidentity}, to simplify the sum of $\ncr{m}{k}$ which emerges as we manipulate the quantities below. We find
\begin{eqnarray}
\E[M] & \ = \ & \sum_{m = k}^N m \frac{\ncr{m-1}{k-1}}{\ncr{N}{k}} \nonumber\\  & \ = \ &  \sum_{m = k}^N m \frac{(m-1)!}{(k-1)!(m-k)!} \frac{k!(N-k)!}{N!}\nonumber\\ &=& \sum_{m=k}^N \frac{m! k}{k!(m-k)!} \frac{k!(N-k)!}{N!}\nonumber\\ &=& \frac{k \cdot k!(N-k)!}{N!} \sum_{m=k}^N \ncr{m}{k} \nonumber\\ &=& \frac{k \cdot k!(N-k)!}{N!} \ncr{N+1}{k+1} \nonumber\\ &=& \frac{k \cdot k!(N-k)!}{N!} \frac{(N+1)!}{(k+1)!(N-k)!} \nonumber\\ &=& \frac{k(N+1)}{k+1}. \nonumber \end{eqnarray}

As we have such a clean expression, it's trivial to solve for $N$ in terms of $k$ and $\E[M]$: $$N \ = \ \E[M] \left(1 + \frac1{k}\right) - 1.$$ Thus if we substitute in $m$ (our observed value for $M$) as our best guess for $\E[M]$, we obtain our estimate for the number of tanks produced: $$\widehat{N} \ = \ m\left(1 + \frac1{k}\right) - 1,$$ completing the proof. \hfill $\Box$

\begin{rek} A more advanced analysis can prove additional results about our estimator, for example, whether or not it is unbiased.\end{rek}

\begin{rek} There are many ways to see this formula is reasonable. The first is to try extreme cases, such as $k=N$ (which forces $m$ to be $N$ and gives $N$ as the answer), or to try $k=1$. In that case we expect our one observation to be around $N/2$, and thus a formula that has the best guess being doubling the observation is logical. We can also get close to this formula from by trying to guess the functional form (for more details see Appendix \ref{sec:gtpregression}). We know our best guess must be at least $m$, so let's write it as $m + f(m,k)$. For a fixed $k$ as $m$ increases we might expect our guess to increase, while for fixed $m$  as $k$ increases we would expect a smaller boost. These heuristics suggest $f(m,k)$ increases with $m$ and decreases with $k$; the simplest such function is $b m/k$ for some constant $b$. This leads to a guess of $m + bm/k$, and again looking at extreme cases we get very close to the correct formula. \end{rek}

%%%%%%%%%%%%%%%%%%%%%%%%%%%%%%%%%%%%%%%%%%%%%%%%%%%%%%%%%%%%%%%%%%%%%%%%%%%%%%%%%%%%%%%%%%%%%%%%%%%%%%%%%%%%%%%%%%%%%%%
%%%%%%%%%%%%%%%%%%%%%%%%%%%%%%%%%%%%%%%%%%%%%%%%%%%%%%%%%%%%%%%%%%%%%%%%%%%%%%%%%%%%%%%%%%%%%%%%%%%%%%%%%%%%%%%%%%%%%%%
%%%%%%%%%%%%%%%%%%%%%%%%%%%%%%%%%%%%%%%%%%%%%%%%%%%%%%%%%%%%%%%%%%%%%%%%%%%%%%%%%%%%%%%%%%%%%%%%%%%%%%%%%%%%%%%%%%%%%%%
\section{Derivation with an unknown minimum}\label{sec:unknownminimum}

Not surprisingly, when we do not know the lowest serial number the resulting algebra becomes more involved; fortunately, though, with a bit of work we are still able to get nice closed-form expressions for the needed sums and obtain again a clean answer for the estimated number of tanks. We still assume the tanks are numbered sequentially, and focus on the spread (the difference between the largest and smallest observed values). Similar to the previous section, we derive a formula to inflate the observed spread to be a good estimate of the number of total tanks.

We first set some notation:

\begin{itemize}
\item the minimum tank serial value, $N_1$,
\item the maximum tank serial value, $N_2$,
\item the total number of tanks, $N$ ($N = N_2 - N_1 + 1$),
\item the observed minimum value, $m_1$ (with corresponding random variable $M_1$),
\item the observed maximum value, $m_2$ (with corresponding random variable $M_2$),
\item the observed spread $s$ (with corresponding random variable $S$).

\end{itemize}

As $s = m_2 - m_1$, in the arguments below we can focus on just $s$ and $S$. We will prove the best guess is $s\left(1 + \frac{2}{k-1}\right) - 1$.

\begin{rek} There are two differences between this formula and the case when the smallest serial number is known. The first is we divide by $k-1$ and not $k$; however, as we cannot estimate a spread with one observation this is reasonable. Note the similarity here with the sample standard deviation, where we divide by one less than the number of observations; while one point suffices to estimate a mean, we need at least two for the variance. The second difference is that we have a factor of 2, which can be interpreted as movement in both directions. \end{rek}

%%%%%%%%%%%%%%%%%%%%%%%%%%%%%%%%%%%%%%%%%%%%%%%%%%%%%%%%%%%%%%
%%%%%%%%%%%%%%%%%%%%%%%%%%%%%%%%%%%%%%%%%%%%%%%%%%%%%%%%%%%%%%
%%%%%%%%%%%%%%%%%%%%%%%%%%%%%%%%%%%%%%%%%%%%%%%%%%%%%%%%%%%%%%
\subsection{The probability that the spread is $s$}

We claim that if we observe $k$ tanks then for $k-1 \le s \le N_2 - N_1$ we have $${\rm Prob}(S = s) \ = \ \frac{\sum_{m = N_1}^{N_2-s}{{s-1}\choose{k-2}}}{{{N_2-N_1+1}\choose{k}}}\ =\ \frac{\left(N_2-N_1+1-s\right){{s-1}\choose{k-2}}}{{{N_2-N_1+1}\choose{k}}} \ = \ \frac{\left(N-s\right){{s-1}\choose{k-2}}}{\ncr{N}{k}},$$ and for all other $s$ the probability is zero.

To see this, note that the spread $s$ must be at least $k-1$ (as we have $k$ observations), and cannot be larger than $N_2-N_1$. If we want a spread of $s$, if the smallest observed value is $m$ then the largest is $m+s$. We must choose exactly $k-2$ of the $s - 1$ numbers in $\{m+1, m+2, \dots, m+s-1\}$; there are $\ncr{s-1}{k-2}$ ways to do so. This proves the first equality, the sum over $m$. As all the summands are the same we get the second equality, and the third follows from our definition of $N$. \hfill $\Box$

\ \\
%%%%%%%%%%%%%%%%%%%%%%%%%%%%%%%%%%%%%%%%%%%%%%%%%%%%%%%%%%%%%%
%%%%%%%%%%%%%%%%%%%%%%%%%%%%%%%%%%%%%%%%%%%%%%%%%%%%%%%%%%%%%%
%%%%%%%%%%%%%%%%%%%%%%%%%%%%%%%%%%%%%%%%%%%%%%%%%%%%%%%%%%%%%%
\subsection{The best guess for $\widehat{N}$}

We argue similarly as in the previous section. In the algebra below we will use our second binomial identity, \eqref{eq:secondcombidentity}; relabeling the parameters it is \begin{eqnarray}\label{eq:secondcombidentitygeneralizednew}\sum_{\ell=a}^{b} \ncr{\ell}{a} \ =\ \ncr{b+1}{a+1}.\end{eqnarray}

We begin by computing the expected value of the spread. We include all the details of the algebra; the idea is to manipulate the expressions and pull out terms that are independent of the summation variable, and rewrite expressions so that we can identify binomial coefficients and then apply our combinatorial results. We have \begin{eqnarray} \E[S] &\ =\ & \sum_{s=k-1}^{N-1} s {\rm Prob}(S=s) \nonumber\\ &=& \sum_{s=k-1}^{N-1} s \frac{\left(N-s\right){{s-1}\choose{k-2}}}{\ncr{N}{k}} \nonumber\\ &=&
 \ncr{N}{k}^{-1} \sum_{s=k-1}^{N-1} s\left(N-s\right) {{s-1}\choose{k-2}} \nonumber\\
& \ = \ &  \ncr{N}{k}^{-1} N \sum_{s=k-1}^{N-1} \frac{s (s-1)!}{(s-k+1)! (k-2)!} - \ncr{N}{k}^{-1} \sum_{s=k-1}^{N-1} \frac{s^2 (s-1)!}{(s-k+1)! (k-2)!}\nonumber\\
& \ = \ &  \ncr{N}{k}^{-1} N \sum_{s=k-1}^{N-1} \frac{s! (k-1)}{(s-k+1)! (k-1)!} - \ncr{N}{k}^{-1} \sum_{s=k-1}^{N-1} \frac{s s! (k-1)}{(s-k+1)! (k-1)!} \ = \ T_1 - T_2. \nonumber
\end{eqnarray}

We first simplify $T_1$; below we always try to multiply by 1 in such a way that we can combine ratios of factorials into binomial coefficients: \begin{eqnarray} T_1 & \  = \ & \ncr{N}{k}^{-1} N \sum_{s=k-1}^{N-1} \frac{s! (k-1)}{(s-k+1)! (k-1)!} \nonumber\\
& = & \ncr{N}{k}^{-1} N(k-1) \sum_{s=k-1}^{N-1} \ncr{s}{k-1} \nonumber\\
&=& \ncr{N}{k}^{-1} N(k-1) \ncr{N}{k} \ = \ N (k-1), \nonumber \end{eqnarray} where we used \eqref{eq:secondcombidentitygeneralizednew} with $a=k-1$ and $b=N-1$.

Turning to $T_2$ we argue similarly, at one point replacing $s$ with $(s-1)+1$ to assist in collecting factors into a binomial coefficient: \begin{eqnarray}
T_2 & \ = \ &  \ncr{N}{k}^{-1} \sum_{s=k-1}^{N-1} \frac{s s! (k-1)}{(s-k+1)! (k-1)!} \nonumber\\
& \ = \ &  \ncr{N}{k}^{-1} \sum_{s=k-1}^{N-1} \frac{(s+1-1) s! (k-1)}{(s-(k-1))! (k-1)!}\nonumber\\
& \ = \ &  \ncr{N}{k}^{-1} \sum_{s=k-1}^{N-1} \frac{(s+1)! (k-1) k}{(s+1-k)! (k-1)! k} - \ncr{N}{k}^{-1} \sum_{s=k-1}^{N-1} \frac{s! (k-1)}{(s-(k-1))! (k-1)!}\nonumber\\
& \ = \ &  \ncr{N}{k}^{-1} \sum_{s=k-1}^{N-1} \frac{(s+1)! k (k-1)}{(s+1-k)! k!} - \ncr{N}{k}^{-1} \sum_{s=k-1}^{N-1} (k-1) \ncr{s}{k-1} \nonumber\\
& \ = \ & \ncr{N}{k}^{-1} \sum_{s=k-1}^{N-1} k(k-1) \ncr{s+1}{k}  - \ncr{N}{k}^{-1} \sum_{s=k-1}^{N-1} (k-1) \ncr{s}{k-1} \ = \ T_{21} + T_{22}. \nonumber
\end{eqnarray}

We can immediately evaluate $T_{22}$ by using \eqref{eq:secondcombidentitygeneralized} with $a=k-1$ and $b=N-1$, and find $$T_{22} \ = \ \ncr{N}{k}^{-1} (k-1) \ncr{N}{k}\ =\ k-1.$$

Thus all that remains is analyzing $T_{21}$: $$T_{21} \ = \ \ncr{N}{k}^{-1} \sum_{s=k-1}^{N-1} \ncr{s+1}{k} k(k-1).$$
We pull $k(k-1)$ outside the sum, and letting $w = s+1$ we see that $$T_{21} \ = \ \ncr{N}{k}^{-1} k(k-1) \sum_{w=k}^{N} \ncr{w}{k}, $$ and then from \eqref{eq:secondcombidentitygeneralized} with $a=k$ and $b=N$ we obtain
$$T_{21} \ = \ \ncr{N}{k}^{-1} k(k-1) \sum_{w=k}^{N} \ncr{w}{k}\ =\ \ncr{N}{k}^{-1} k(k-1) \ncr{N+1}{k+1}.$$

Thus substituting everything back yields $$\E[S] \ = \  N(k-1) + (k-1) - \ncr{N}{k}^{-1} k(k-1) \ncr{N+1}{k+1}.$$ We can simplify the right hand side:
\begin{eqnarray}
(N+1)(k-1)-k(k-1) \frac{\frac{(N+1)!}{(N-k)! (k+1)!}}{\frac{N!}{(N-k)! k!}} & \ = \ &  (N+1)(k-1)-k(k-1) \frac{(N+1)! (N-k)! k!}{N! (N-k)! (k+1)!}\nonumber\\
& \ = \ &  (N+1)(k-1)-k(k-1) \frac{N+1}{k+1}\nonumber\\
& \ = \ &  (N+1)(k-1) - \frac{k(k-1) (N+1)}{k+1}\nonumber\\
& \ = \ &  (N+1)(k-1)\left(1-\frac{k}{k+1}\right)\nonumber\\
& \ = \ &  (N+1) \frac{k-1}{k+1}, \nonumber
\end{eqnarray}
and thus obtain $$\E[S]\ = \ (N+1) \frac{k-1}{k+1}.$$

The analysis is completed as before, where we pass from our observation of $s$ for $S$ to a prediction $\widehat{N}$ for $N$: $$\widehat{N} \ = \ \frac{k+1}{k-1} s - 1 \ = \ s\left(1 + \frac{2}{k-1}\right) - 1,$$ where the final equality is due to rewriting the algebra to mirror more closely the formula from the case where the first tank is numbered 1. Note that this formula passes the same smell checks the other did; for example $s \frac{2}{k-1} - 1$ is always at least 1 (remember $k$ is at least 2), and thus the lowest estimate we can get for the number of tanks is $s+1$.

%%%%%%%%%%%%%%%%%%%%%%%%%%%%%%%%%%%%%%%%%%%%%%%%%%%%%%%%%%%%%%%%%%%%%%%%%%%%%%%%%%%%%%%%%%%%%%%%%%%%%%%%%%%%%%%%%%%%%%%
%%%%%%%%%%%%%%%%%%%%%%%%%%%%%%%%%%%%%%%%%%%%%%%%%%%%%%%%%%%%%%%%%%%%%%%%%%%%%%%%%%%%%%%%%%%%%%%%%%%%%%%%%%%%%%%%%%%%%%%
%%%%%%%%%%%%%%%%%%%%%%%%%%%%%%%%%%%%%%%%%%%%%%%%%%%%%%%%%%%%%%%%%%%%%%%%%%%%%%%%%%%%%%%%%%%%%%%%%%%%%%%%%%%%%%%%%%%%%%%
\section{Comparison of Approaches}

\begin{figure}[h]
\begin{center}
\scalebox{1}{\includegraphics{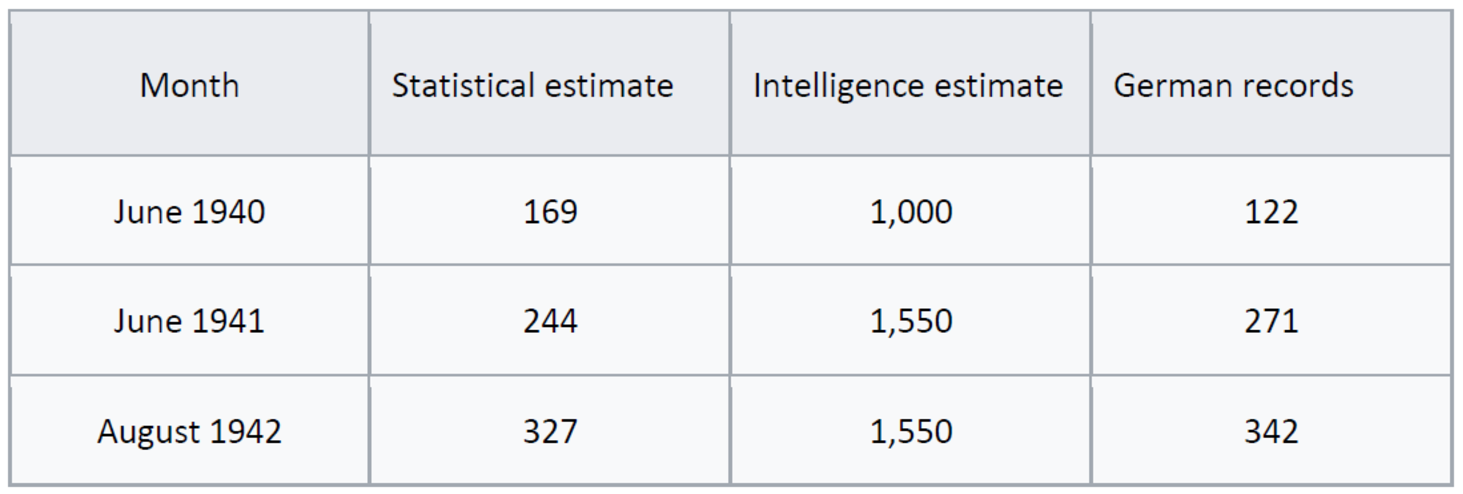}}\caption{\label{fig:spiesversusstats} Comparison of estimates from statistics and spies to the true values. Table from \cite{A5}.}
\end{center}\end{figure}

So, which did better: statistics or spies? Once the Allies won the war, they could look into Albert Speer's, the Nazi Minister of Armaments, records to see the exact number of tanks produced each month; see Figure \ref{fig:spiesversusstats}.

The meticulous German record keeping comes in handy for the vindication of the statisticians; these estimates were astoundingly more accurate. While certainly not perfect (an underestimation of 30 tanks could have pretty dire consequences when high command is allocating resources), the statistical analysis was tremendously superior to the intelligence estimates, which were off by factors of 5 or more. We mentioned earlier the lessons to be learned from McClellan's caution. He was the first General of the Army of the Potomac (which was the Union army headquartered near Washington), and he repeatedly missed opportunities to deliver a debilitating blow to General Robert E. Lee's army of Northern Virginia, most famously during Lee's retreat from Antietam. Despite vastly outnumbering Lee in men and supplies, McClellan chronically overestimated Lee's forces, causing him to be overly cautious and far too timid a commander. Ultimately the Civil War would drag on for four years and costing over 650,000 American lives, and one wonders how the outcome would have been different if McClellan had been more willing to take the field.

We encourage the reader to write some simple code to simulate both problems discussed here (or see \cite{A4}), namely when we know and when we don't know the number of the lowest tank. These problems provide a valuable warning on how easy it is to accidentally convey information. In many situations today numbers are randomly generated to prevent such an analysis. Alternatively, sometimes numbers are deliberately started higher to fool an observer into thinking that more is present than actually is (examples frequently seen are the counting done during a workout, putting money in the tip jar at the start of the shift to encourage future patrons to be generous, or checkbooks starting with the first check as 100 or higher so the recipient does not believe it is from a new account).

\appendix
%%%%%%%%%%%%%%%%%%%%%%%%%%%%%%%%%%%%%%%%%%%%%%%%%%%%%%%%%%%%%%%%%%%%%%%%%%%%%%%%%%%%%%%%%%%%%%%%%%%%%%%%%%%%%%%%%%%%%%%
%%%%%%%%%%%%%%%%%%%%%%%%%%%%%%%%%%%%%%%%%%%%%%%%%%%%%%%%%%%%%%%%%%%%%%%%%%%%%%%%%%%%%%%%%%%%%%%%%%%%%%%%%%%%%%%%%%%%%%%
%%%%%%%%%%%%%%%%%%%%%%%%%%%%%%%%%%%%%%%%%%%%%%%%%%%%%%%%%%%%%%%%%%%%%%%%%%%%%%%%%%%%%%%%%%%%%%%%%%%%%%%%%%%%%%%%%%%%%%%
\section{The German Tank Problem and Linear Regression}\label{sec:gtpregression}

The German Tank Problem is frequently used in probability or discrete math classes, as it illustrates the power of those two disciplines to use binomial identities to great advantage. It's also seen in statistics classes in discussing how to find good estimators of population values. Focusing on these examples, however, neglects another great setting where it may be used effectively: as an application of the power of Linear Regression (or the Method of Least Squares). We quickly review how these methods yield the best-fit line or hyperplane, and then generalize to certain non-linear relationships. We show how simulations can be used to provide support for formulas. This is extremely important, as often we are unable to prove conjectured relationships. Returning to World War II, the Allies could run trials (say drawing numbered pieces of paper from a bag) to model the real world problem, and use the gathered data to sniff out the relationship $m$, $k$ and $N$.

Additionally, we use this appendix as an opportunity to discuss some of the issues that can arise when implementing the Method of Least Squares to find the best fit line. While these do not occur in most applications, it is worth knowing that they can happen and seeing solutions.

%%%%%%%%%%%%%%%%%%%%%%%%%%%%%%%%%%%%%%%%%%%%%%%
%%%%%%%%%%%%%%%%%%%%%%%%%%%%%%%%%%%%%%%%%%%%%%%
%%%%%%%%%%%%%%%%%%%%%%%%%%%%%%%%%%%%%%%%%%%%%%%
\subsection{Theory of Regression}

Suppose we believe there are choices of $a$ and $b$ such that given an input $x$ we should observe $y = ax + b$, but we don't know what these values are. We could observe a large number of pairs of data $\{x_i, y_i\}_{i=1}^I$, and use these to find the values of $a$ and $b$ that minimize the sum of the squares of the errors\footnote{We cannot just add the errors, as then a positive error could cancel with a negative error. We could take the sum of the absolute values, but the absolute value function is not differentiable; it is to have calculus available that we measure errors by sums of squares.} between the observed and predicted values: $$E(a,b) \ = \ \sum_{i=1}^I \left(y_i - (a x_i + b)\right)^2.$$ By setting $$\frac{\partial E}{\partial a} \ = \ \frac{\partial E}{\partial b} \ = \ 0,$$ after some algebra\footnote{The resulting matrix is invertible, and hence there is a unique solution, so long as at least two of the $x_i$'s differ. One can see this through some algebra, where the determinant of the matrix is essentially the variance of the $x_i$'s; if they are not all equal then the variance is positive. If the $x_i$'s are all equal the inverse does not exist, but in such a case we should not be able to predict how $y$ varies with $x$ as we are not varying $x$!} we find the best fit values are $$\vectwo{\widehat{a}}{\widehat{b}} \ = \ \mattwo{\sum_{i=1}^I x_i^2}{\sum_{i=1}^I x_i}{\sum_{i=1}^I x_i}{\sum_{i=1}^I 1}^{-1} \vectwo{\sum_{i=1}^I x_i y_i}{\sum_{i=1}^I y_i};$$ see for example the supplemental material online for \cite{A1}. What matters is that the relation is linear in the unknown parameters $a$ and $b$ (or more generally $a_1, \dots, a_\ell$); similar formulas hold for $$y \ = \ a_1 f_1(x) + \cdots + a_\ell f(x_\ell).$$ For a linearly algebraic approach to regression see for example \cite{A2}.

Regression is a rich subject; we wish to try to find the best fit parameters to relate $N$ to $m$ and $k$; however, we'll shortly see that our initial guess at a relationship is non-linear. Fortunately, by taking logarithms, we can convert many non-linear relations to linear ones, and thus the formulas above are available again. The idea is that by doing extensive simulations we can gather enough data to make a good conjecture on the relationship. Sometimes, as will be the case with the German Tank Problem, we are able to do a phenomenal job in predicting the functional form and coefficients, while other times we can only get some values with confidence.

\begin{figure}[h]
\begin{center}
\scalebox{.7}{\includegraphics{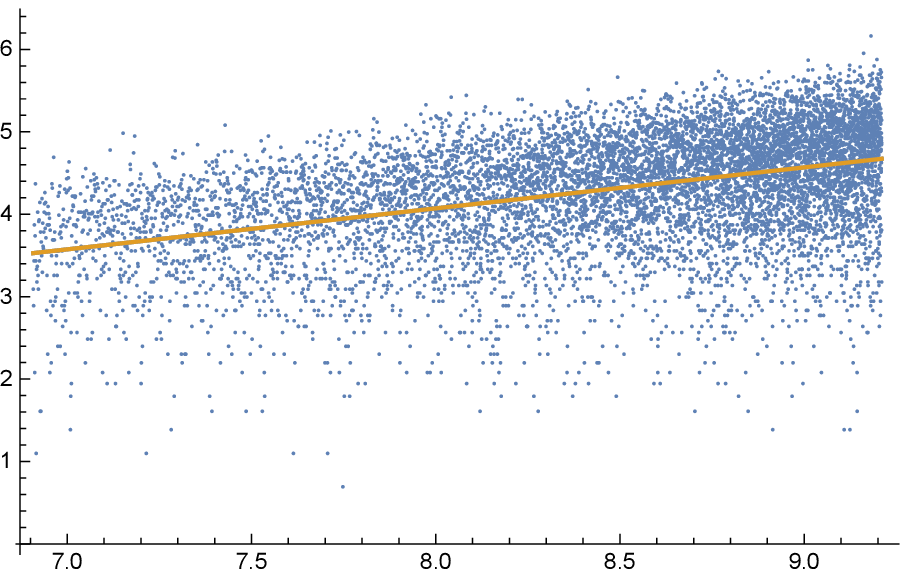}}\ \ \scalebox{.7}{\includegraphics{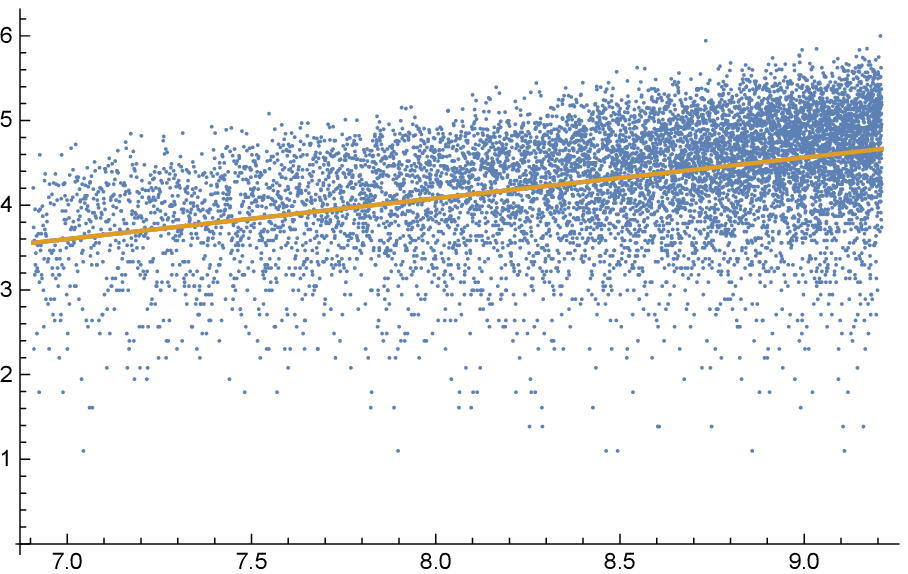}}\ \
\caption{\label{fig:bday10000sims10000to100000sim} Plot of best fit line for $P$ as a function of $D$. We twice ran 10,000 simulations with $D$ chosen from $10,000$  to $100,000$. Best fit values were $a \approx 0.506167, b \approx -0.0110081$ (left) and $a \approx 0.48141$, $b \approx 0.230735$ (right).}
\end{center}\end{figure}

To highlight these features we first quickly review a well-known problem: The Birthday Paradox (see for example \cite{A1}). The standard formulation assumes we have a year with $D$ days, and asks how many people do we need in a room to have a 50\% chance that at least two share a birthday, under the assumption that the birthdays are independent and uniformly distributed from 1 to $D$. A straightforward analysis shows the answer is approximately $D^{1/2} \sqrt{\log 4}$. We now consider the closely related but less well-known problem of what is the expected number of people $P$ we need in a room before there is a match.\footnote{As a nice exercise, use linearity of expectation to show that we expect at least two people to share a birthday when $P = D^{1/2} \sqrt{2} + 1$.} Based on the first problem it is reasonable to expect the answer to also be on the order of $D^{1/2}$, but what is the constant factor? We can try a relation of the form $P = B D^a$, and then taking logs (and setting $b = \log B$) we would get $\log P = a \log D + b$. See Figure \ref{fig:bday10000sims10000to100000sim}.

The two simulations both have similar values for $a$, with both of them consistent with an exponent of 1/2. Unfortunately the values for $b$ wildly differ, though of the two parameters we care more about $a$ as it tells us how our answer changes with the number of days. There is an important lesson here: data analysis can often suggest much of the answer, but it is not always the full story and there is a role for theory in supplementing such analysis.

%%%%%%%%%%%%%%%%%%%%%%%%%%%%%%%%%%%%%%%%%%%%%%%
%%%%%%%%%%%%%%%%%%%%%%%%%%%%%%%%%%%%%%%%%%%%%%%
%%%%%%%%%%%%%%%%%%%%%%%%%%%%%%%%%%%%%%%%%%%%%%%
\subsection{Issues in Applying to the German Tank Problem}

\begin{figure}
\begin{center}
\scalebox{1}{\includegraphics{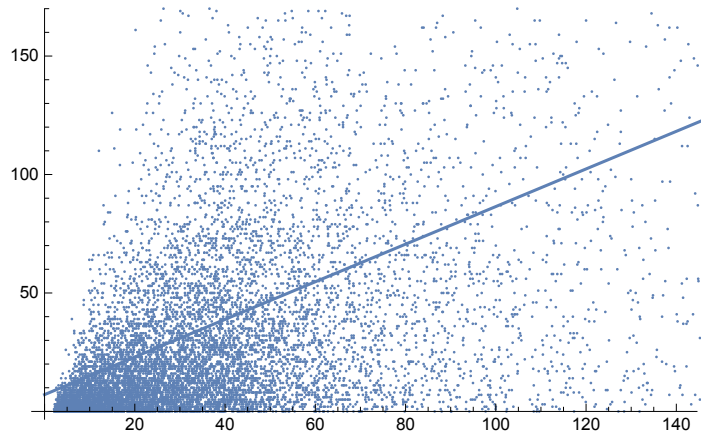}}\caption{\label{fig:listplot1000Nminusmvsmk} Plot of best fit line for $N - m$ as a function of $m/k$. We ran 10,000 simulations with $N$ chosen from $[100, 2000]$ and $k$ from $[10,50]$. Best fit values for $N-m = a(m/k) + b$ for this simulation were $a \approx 0.793716$, $b \approx 7.10602$.}
\end{center}\end{figure}

Building on this lesson, we return to the German Tank Problem. What is a reasonable choice for $N$ as a function of $m$ and $k$? Clearly $N$ is at least $m$, so we try $N = m + f(m,k)$, which transfers the problem to estimating $f(m,k)$. We expect that as $m$ increases this should increase, and as $k$ increases it should decrease. Looking at extreme cases is useful; if $k=N$ then $f(N,N)$ should vanish, as then $m$ must equal $N$. The simplest function that fits this is $f(m,k) = b \cdot m/k$ with $b$ as our free parameter, and we are led to conjecturing a relationship of the form $$N \ =  \ m + b \frac{m}{k} \ = \ m \left(1 + \frac{b}{k}\right).$$ Note that this guess is quite close to the true answer, but because the observed quantities $m$ and $k$ appear as they do, it is not a standard regression problem. We could try to fix this by looking at $N - m$, the number of tanks we need to add to our observed largest value to get the true number. We could then try to write this as a linear function of the ratio $m/k$: $$N - m \ = \ a \frac{m}{k} + b,$$ where we allowed ourselves a constant term to increase our flexibility of what we can model. While for $a=-b=1$ this reproduces the correct formula, finding the best fit values leads to a terrible fit, as evidenced in Figure \ref{fig:listplot1000Nminusmvsmk}.

Why is the agreement so poor, given that proper choices exist? The problem is the way $m$ and $k$ interact, and in the set-up above we have the observed quantity $m$ both as an input variable and as an output in the relation. We thus need a way to separate $m$ and $k$, keeping both on the input side. As remarked, we can do this through logarithms; we discuss another approach in the next subsection.

%%%%%%%%%%%%%%%%%%%%%%%%%%%%%%%%%%%%%%%%%%%%%%%
%%%%%%%%%%%%%%%%%%%%%%%%%%%%%%%%%%%%%%%%%%%%%%%
%%%%%%%%%%%%%%%%%%%%%%%%%%%%%%%%%%%%%%%%%%%%%%%
\subsection{Resolving Implementation Issues}

\begin{figure}[h]
\begin{center}
\scalebox{.6}{\includegraphics{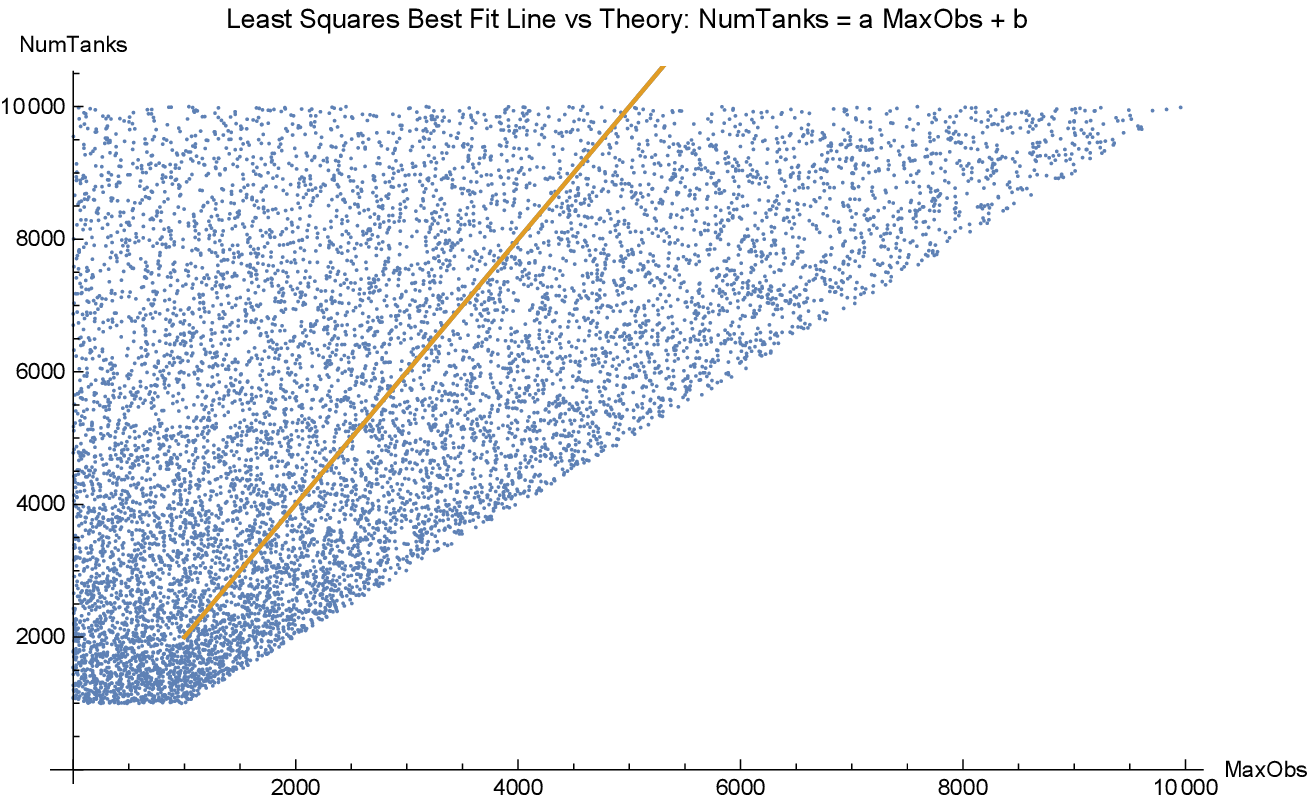}}\ \ \scalebox{.6}{\includegraphics{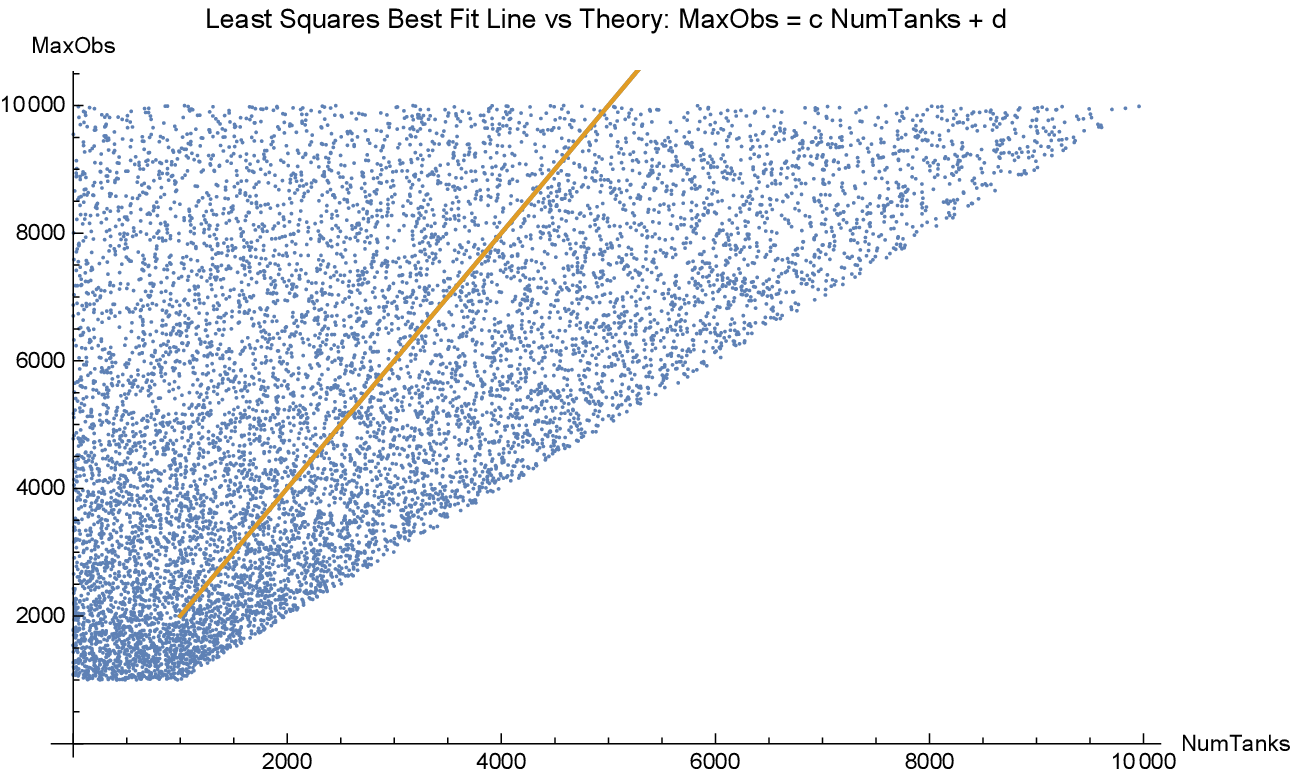}}\caption{Left: Plot of $N$ vs maximum observed tank $m$ for fixed $k=1$. Theory: $N = 2m - 1$, best fit $N = .784m +2875$. Right: Plot of maximum observed tank $m$ vs $N$ for fixed $k=1$. Theory: $m = .5 N + .5$, best fit $m = .496N + 10.5$.}\label{fig:lsfkis1}
\end{center}\end{figure}

\begin{figure}[h]
\begin{center}
\scalebox{.5}{\includegraphics{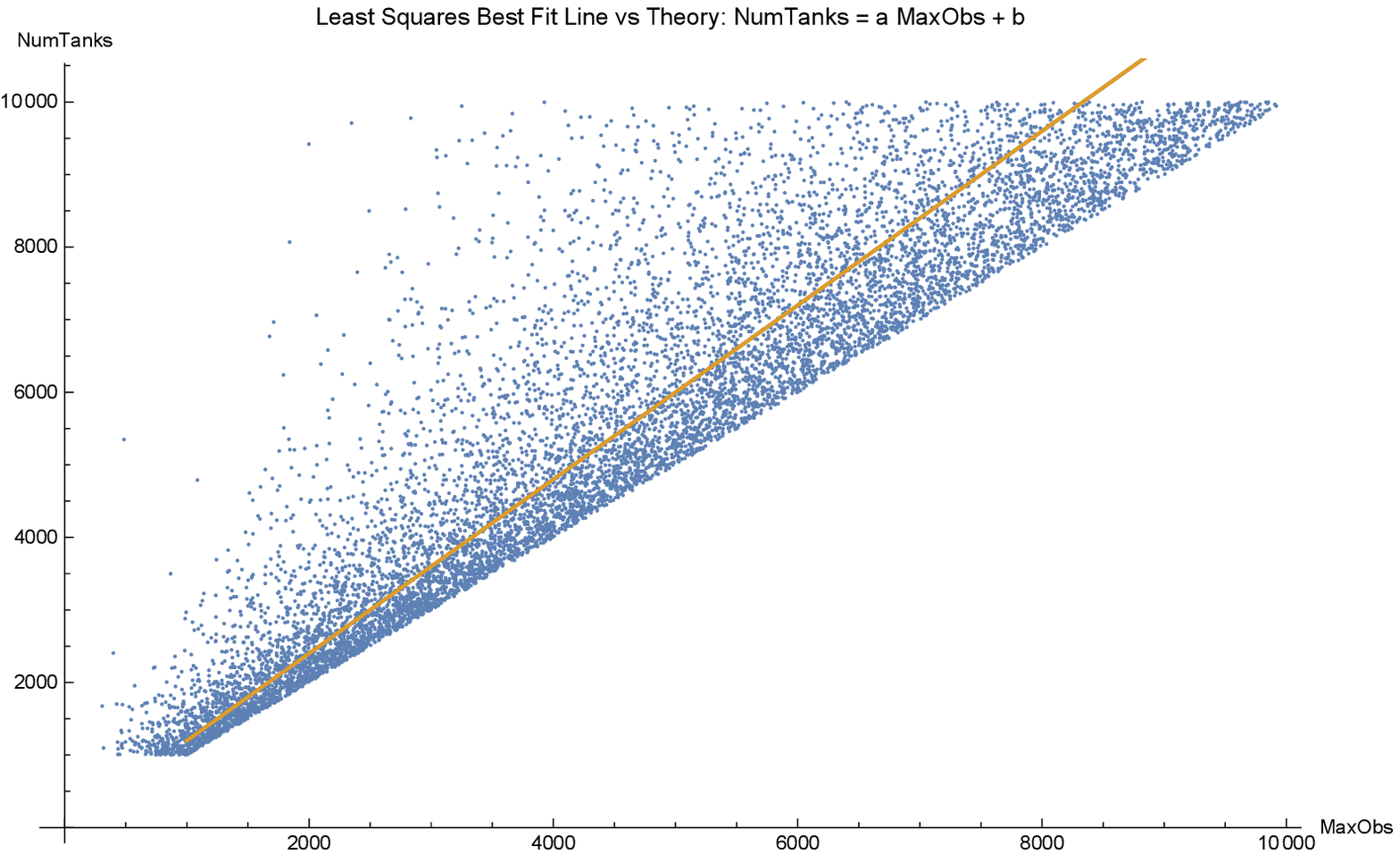}} \ \ \scalebox{.5}{\includegraphics{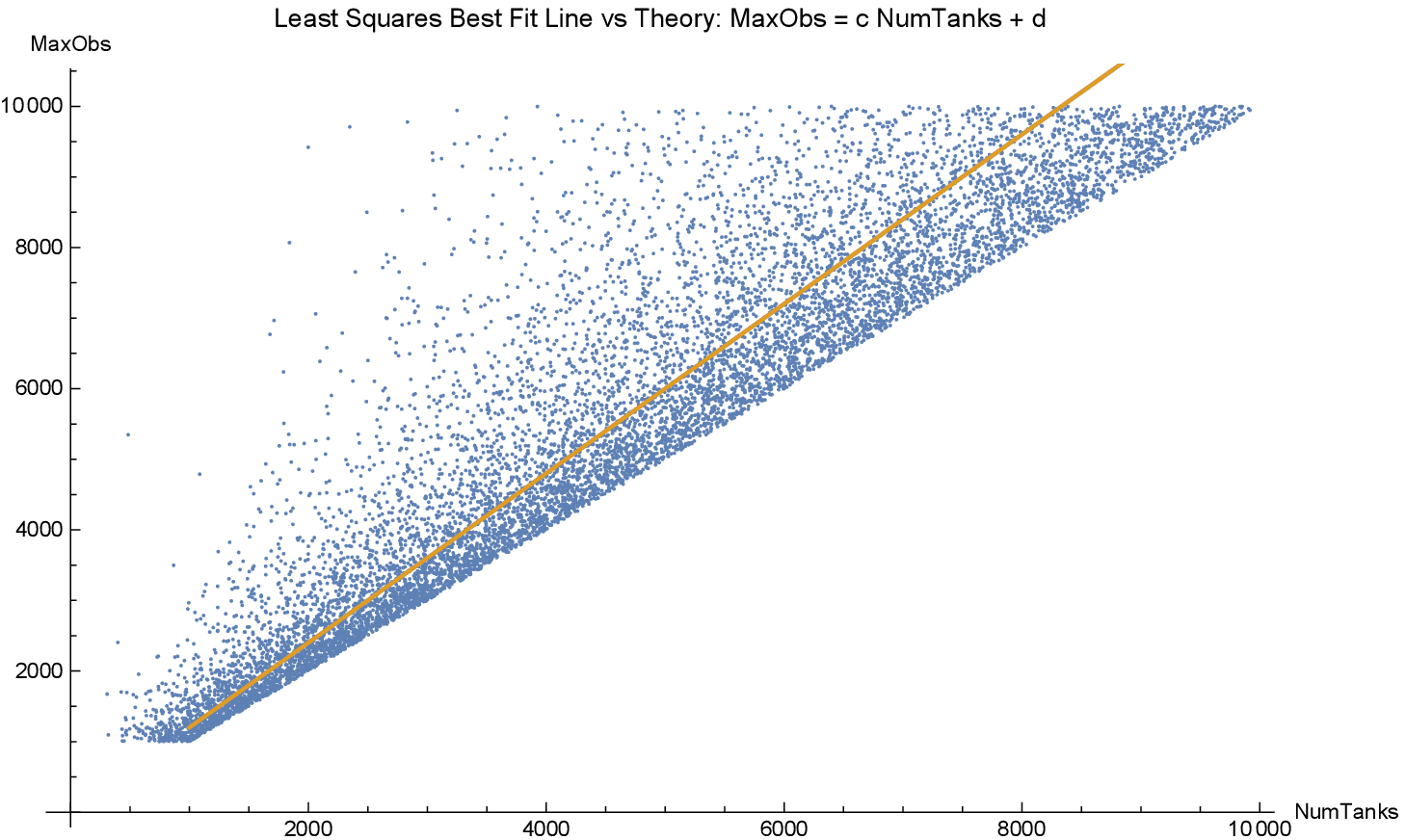}}\caption{Left: Plot of $N$ vs maximum observed tank $m$ for fixed $k=5$. Theory: $N = 1.2m - 1$, best fit $N = 1.037m +749$. Right: Plot of maximum observed tank $m$ vs $N$ for fixed $k=5$. Theory: $m = .883 N + .883$, best fit $m = .828N + 25.8$.}\label{fig:lsfkis5}
\end{center}\end{figure}

We look at our best fit line for two choices of $k$; The left side of Figure \ref{fig:lsfkis1} does $k=1$ while Figure \ref{fig:lsfkis5} is $k=5$. Both of these show a terrible fit of $N$ as a linear function of $m$ (for a fixed $k$). In particular, when $k=1$ we expect $N$ to be $2m-1$ but our best fit line is about $.784m +2875$; this is absurd as for large $m$ we predict $N$ to be less than $m$!  Note, however, the situation is completely different if instead we plot $m$ against $N$ (the right hand side of those figures). Clearly if $N$ linearly depends on $m$ then $m$ linearly depends on $N$. When we do the fits this way, the results are excellent.

Note that from the point of view of an experiment, it makes more sense to plot $m$ as the dependent variable and $N$ as the independent, input variable. The reason is the way we simulate; we fix a $k$ and an $N$ and then choose $k$ distinct numbers uniformly from $\{1, \dots, N\}$.

\begin{figure}[h]
\begin{center}
\scalebox{.75}{\includegraphics{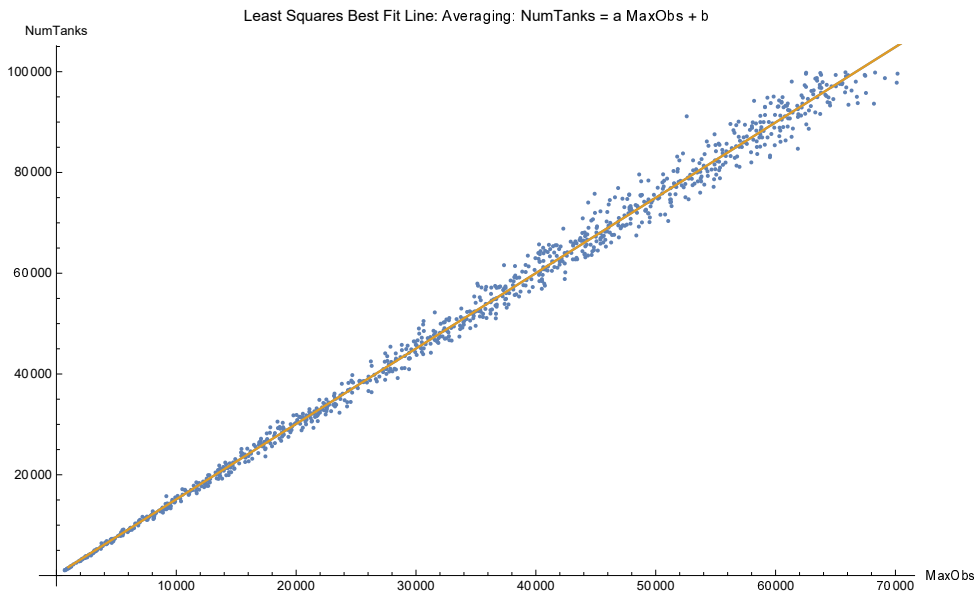}}\caption{Plot of $N$ vs maximum observed tank $m$ for fixed $k=1$. Theory: $N = 1.5 m - 1$, best fit $N = 1.496m + 171.2$.}\label{fig:avekNm}
\end{center}\end{figure}

We end with another approach which works well, and allows us to view $N$ as a function of $m$. Instead of plotting each pair $(m,N)$ for a fixed $k$, we instead fix $k$, choose an $N$, and then do 100 trials. For each trial we record the largest serial number $m$, and then we average these, and plot $(\overline{m}, N)$ where $\overline{m}$ is the average. This greatly decreases the variability, and we now obtain a nearly perfect straight line and fit; see Figure \ref{fig:avekNm}.

%%%%%%%%%%%%%%%%%%%%%%%%%%%%%%%%%%%%%%%%%%%%%%%
%%%%%%%%%%%%%%%%%%%%%%%%%%%%%%%%%%%%%%%%%%%%%%%
%%%%%%%%%%%%%%%%%%%%%%%%%%%%%%%%%%%%%%%%%%%%%%%
\subsection{Determining the Functional Form}

\begin{figure}[h]
\begin{center}
\scalebox{1}{\includegraphics{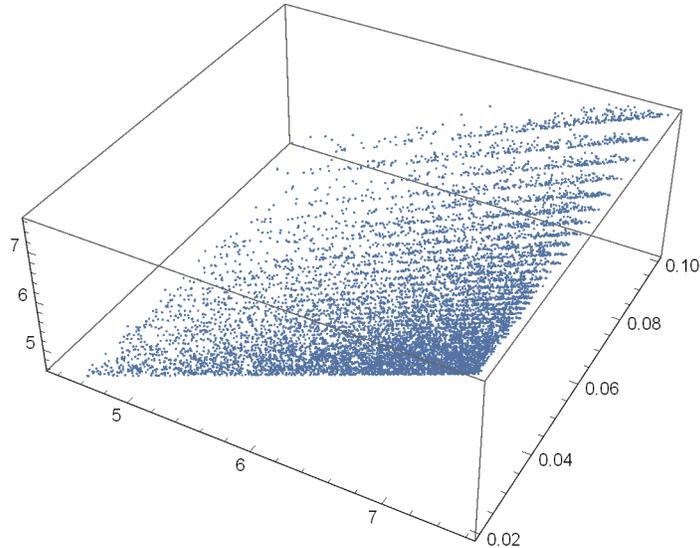}}\caption{\label{fig:listplot10000logmk} Plot of $\log N$ against $\log m$ and $1/k$. We ran 10,000 simulations with $N$ chosen from $[100, 2000]$ and $k$ from $[10,50]$. The data is well-approximated by a plane (we do not draw it in order to prevent our image from being too cluttered).}
\end{center}\end{figure}

We consider the more general relation $$N \ = \ C m^a \left(1 + \frac{b}{k}\right), $$ where we expect $C=a=b=1$; note this won't have the $-1$ summand we know should be there, but for large $m$ that should have negligible impact. Letting $C = e^c$ for notational convenience, we find $$\log N \ = \ c + a \log(m) + \log\left(1 + \frac{b}{k}\right).$$ If $x$ is large then $\log(1 + 1/x) \approx 1/x$, so we try the approximation $$\log N \ \approx \ c + a \log(m) + b \frac1{k}.$$ Figure \ref{fig:listplot10000logmk} shows the analysis when $C=1$ (so $c=0$), as the analysis then reduces to the usual case with two unknown parameters. We chose to take $C=1$ from the lesson we learned in the analysis of the Birthday Problem.

The best fit values of the parameters are $a = 0.999911$ and $b = 0.961167$, which are reasonably close to $a=b=1$. Thus these numerics strongly support our conjectured relation $N = m(1 + 1/k)$, and shows the power of statistics. While we were able to see the arguments needed to prove this relation essentially holds, imagine we could not prove it but still have our heuristic arguments and analysis of extreme cases which suggest it is true. By simulating data and running the regression, we see that our formula does a stupendous job explaining our observations, and thus gain confidence to use it in the field.

\begin{figure}[h]
\begin{center}
\scalebox{.525}{\includegraphics{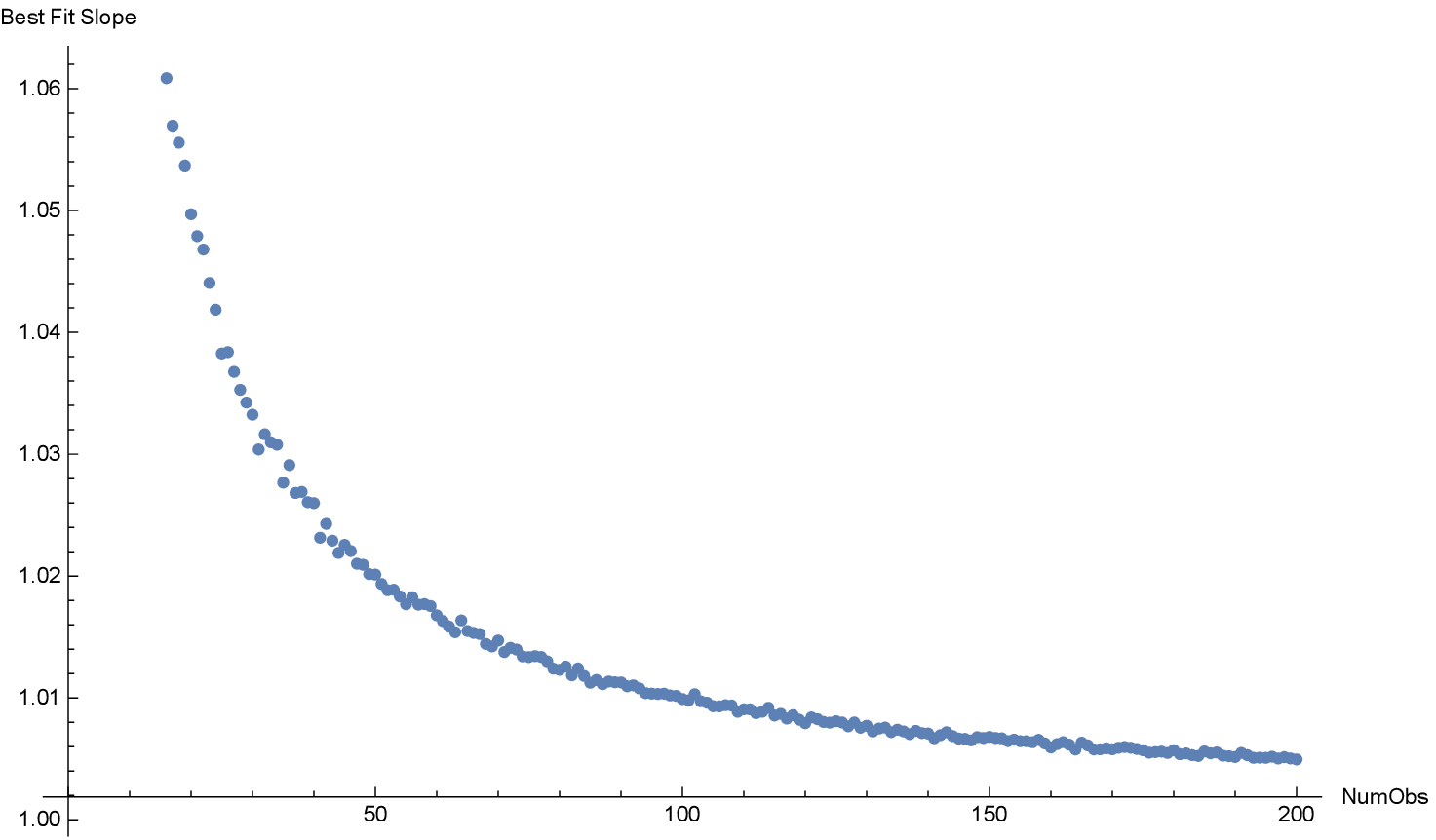}} \ \ \ \scalebox{.525}{\includegraphics{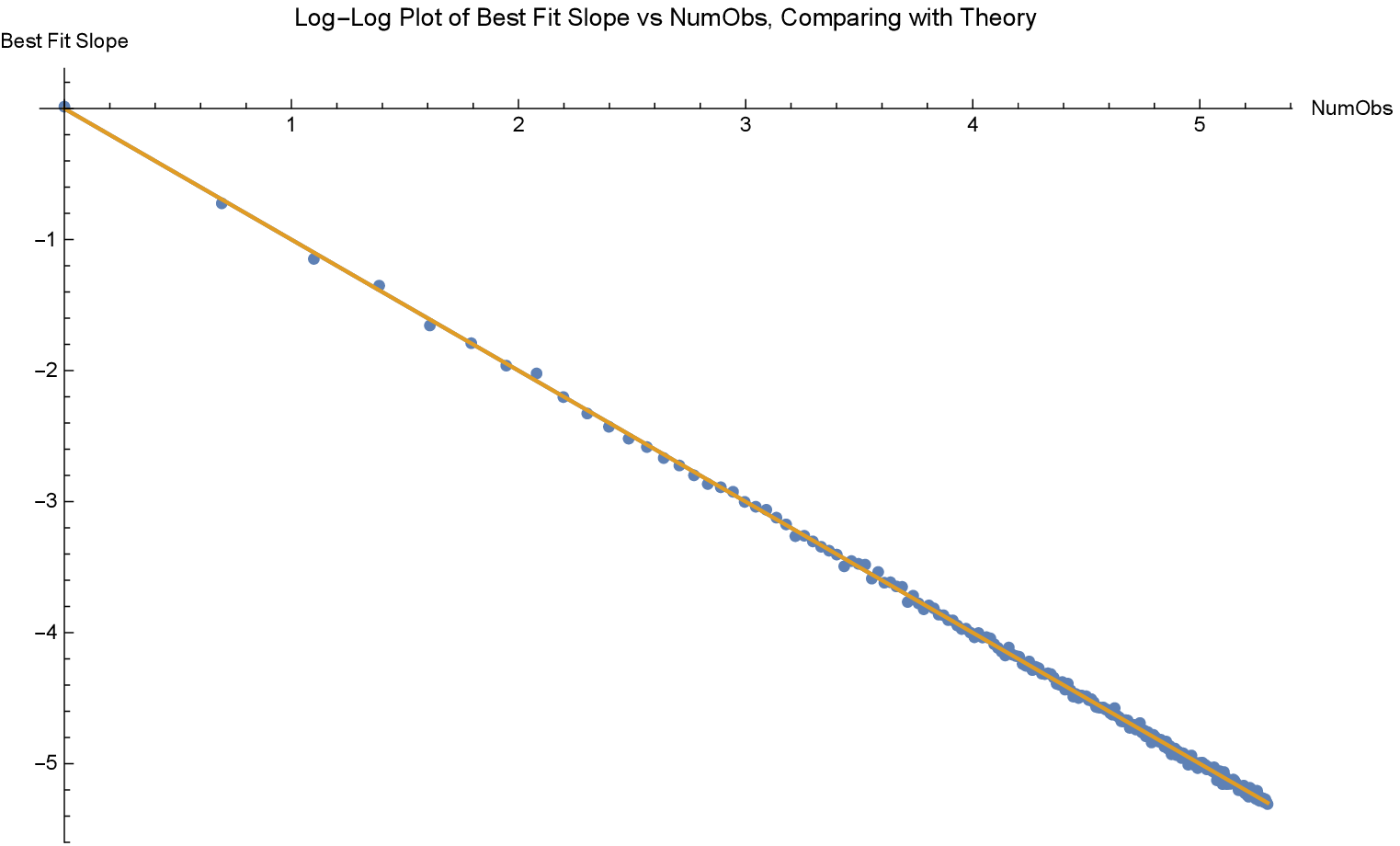}}\caption{Left: Plot of $a(k)$, the slope in $N = a(k) m + b$, versus $k$. Right: Log-Log Plot of $a(k)-1$ versus $k$. In $\log(a(k)-1)$ versus $\log k$, theory is $\log(a(k)-1) = -1 \log(k)$, best fit line is $\log(a(k)-1) = -.999 \log(k) -.007$.}\label{fig:sniffoutk}
\end{center}\end{figure}

We end with one last approach. Let us guess a relationship of the form $N = a(k)m + b(k)$, where $a(k) = 1 + f(k)$ (we write $a(k)$ as $1 + f(k)$ as we know there have to be at least $m$ tanks). We can fix $k$, and find the best fit values of $a(k)$ and $b(k)$. In Figure \ref{fig:sniffoutk} we plot the best fit slope $a(k)$ versus $k$, as well as a log-log plot. For the log-log plot we look at $a(k)-1$, subtracting the known component. We see a beautiful linear relation, and thus even if we did not know it should be $m$ plus a constant times $m/k$, the data suggests that beautifully! Specifically, we found the best fit line was $\log(a(k)-1) = -.999 \log(k) -.007$, suggesting that $a(k) = 1 + 1/k$; we obtain the correction functional form just by running simulations!

%%%%%%%%%%%%%%%%%%%%%%%%%%%%%%%%%%%%%%%%%%%%%%%%%%%%%%%%%%%%%%%%%%%%%%%%%%%%%%%%%%%%%%%%%%%%%%%%%%%%%%%%%%%%%%%%%%%%%%%
%%%%%%%%%%%%%%%%%%%%%%%%%%%%%%%%%%%%%%%%%%%%%%%%%%%%%%%%%%%%%%%%%%%%%%%%%%%%%%%%%%%%%%%%%%%%%%%%%%%%%%%%%%%%%%%%%%%%%%%
%%%%%%%%%%%%%%%%%%%%%%%%%%%%%%%%%%%%%%%%%%%%%%%%%%%%%%%%%%%%%%%%%%%%%%%%%%%%%%%%%%%%%%%%%%%%%%%%%%%%%%%%%%%%%%%%%%%%%%%
%\section{Bibliography}

\ \\


\begin{thebibliography}{99} % '2nd argument contains the widest acronym'

\bibitem{A0}
M. Cozzens and S. J. Miller, \emph{The Mathematics of Encryption: An Elementary Introduction}, AMS Mathematical World series \textbf{29}, Providence, RI, 2013.

\bibitem{A1}
S. J. Miller, \emph{The Probability Lifesaver}, Princeton University Press, Princeton, NJ, 2018. \bburl{https://web.williams.edu/Mathematics/sjmiller/public_html/probabilitylifesaver/index.htm}.

\bibitem{A2}
G. Strang, \emph{Introduction to Linear Algebra}, Fifth Edition, Wellesley-Cambridge Press, 2016.

\bibitem{A3}
Probability and Statistics Blog,  \emph{How many tanks? MC testing the GTP}, \bburl{https://statisticsblog.com/2010/05/25/how-many-tanks-gtp-gets-put-to-the-test/}.

\bibitem{A4}
Statistical Consultants Ltd, \emph{The German Tank Problem}, \bburl{https://www.statisticalconsultants.co.nz/blog/the-german-tank-problem.html}.

\bibitem{A5}
WikiEducator, \emph{Point Estimation - German Tank Problem}, \bburl{https://wikieducator.org/Point_estimation_-_German_tank_problem}.

\bibitem{A6}
Wikipedia, \emph{German Tank Problem}, Wikimedia Foundation, \bburl{https://en.wikipedia.org/wiki/German_tank_problem}.



\end{thebibliography}
\end{document}